\documentclass[aoas,preprint]{imsart}
\setattribute{journal}{name}{} % Remove for Submission

\RequirePackage[OT1]{fontenc}
\RequirePackage{amsthm,amsmath}
\RequirePackage{natbib}
\RequirePackage[colorlinks,citecolor=blue,urlcolor=blue]{hyperref}
\usepackage[pdftex]{graphicx}

\startlocaldefs
\numberwithin{equation}{section}
\theoremstyle{plain}

\DeclareMathOperator*{\argmin}{arg\,min}
\DeclareMathOperator*{\argmax}{arg\,max}
\newcommand{\pkg}[1]{{\fontseries{b}\selectfont #1}}
\endlocaldefs

\begin{document}

\begin{frontmatter}
\title{The Bayesian Sorting Hat: A Decision-theoretic Approach to Size-Constrained Clustering}
\runtitle{The Bayesian Sorting Hat}
%\thankstext{T1}{Footnote to the title with the ``thankstext'' command.}

\begin{aug}
\author{\fnms{Justin D.} \snm{Silverman}\thanksref{m1}\ead[label=e1]{Justin.Silverman@duke.edu}}
\and
\author{\fnms{Rachel} \snm{Silverman}\thanksref{m2}\ead[label=e2]{kloss@live.unc.edu}}

\affiliation{\thanksmark{m1}Duke University and \thanksmark{m2}University of North Carolina, Chapel Hill}

\address{\thanksref{m1} Duke University\\
Center for Genomic and Computational Biology \\
101 Science Drive \\
Durham, NC 27708 \\
\printead{e1}\\
\phantom{E-mail:\ }}

\address{\thanksref{m2} University of North Carolina, Chapel Hill \\
Department of Biostatistics\\
135 Dauer Drive\\
Chapel Hill, NC 27599\\
\printead{e2}}
\end{aug}

\begin{abstract}
Size-constrained clustering (SCC) refers to the dual problem of using observations to determine latent cluster structure while at the same time assigning observations to the unknown clusters subject to an analyst defined constraint on cluster sizes. While several approaches have been proposed, SCC remains a difficult problem due to the combinatorial dependency between observations introduced by the size-constraints. Here we reformulate SCC as a decision problem and introduce a novel loss function to capture various types of size constraints. As opposed to prior work, our approach is uniquely suited to situations in which size constraints reflect and external limitation or desire rather than an internal feature of the data generation process. To demonstrate our approach, we develop a Bayesian mixture model for clustering respondents using both simulated and real categorical survey data. Our motivation for the development of this decision theoretic approach to SCC was to determine optimal team assignments for a Harry Potter themed scavenger hunt based on categorical survey data from participants. 
\end{abstract}

% \begin{keyword}
% \kwd{sample}
% \kwd{\LaTeXe}
% \end{keyword}

\end{frontmatter}

\section{Introduction}

Consider the problem of picking teams for a Harry Potter \citep{Rowling1999} themed scavenger hunt based on data from a survey administered to each participant. The teams are to be identified with the four houses in the Harry Potter books (Ravenclaw, Slytherin, Hufflepuff, and Gryffindor) and the method by which these teams are picked should take into account prior knowledge regarding the characteristics of a prototypical member of each of the four houses. In addition, we assert that well picked teams should be of relatively equal size and cluster together individuals based on similar survey responses. This is a typical example of a size-constrained clustering (SCC)  where clustering of observations must account for constraints on the relative size of clusters in addition to identifying latent cluster structure in data. Beyond its application to party planning, SCC is relevant in a variety of other areas including adaptive clinical trial designs where treatment groups are blinded but aggregate totals are known \citep{Teel2015}, in market research when inferring individual purchasing habits given relevant covariate and aggregate purchasing patterns for a region, in the inference of voting patterns given district totals and relevant covariates \citep{Teel2015}, or in various industry applications where resources or supply is limited with some flexibility in its allocation.

Unfortunately, in contrast to traditional clustering problems where observations are typically viewed as independently generated from the latent clusters, SCC is notoriously difficult due to the combinatorial dependency structure introduced by the size constraint. For example, in the extreme case where cluster totals are fixed, changing the cluster assignment of one observations requires that at least one other observation must be reallocated to make room for the reassignment. Prior work on SCC has been limited due to the difficulties of working with such combinatorial constraints. The majority of methods for SCC are based on heuristic algorithms such as the integer linear programming approach of \cite{Zhu2010}, the model-based iterative bipartitioning heuristic of \cite{Zhong2003, Zhong2003a}, the K-Means based methods of \cite{Usami2014} and \cite{Ganganath2014}, the spectral clustering methods of \cite{Chen2006}, and the Parallel Balanced Team Formation Algorithm of \cite{Kim2015}. A EM Estimation method based on the conditional Bernoulli distribution was proposed by \cite{Teel2015}, although this method can handle problems with no more than two latent clusters. In addition, \cite{Park2015} proposed a Bayesian approach for uncovering latent cluster membership in multivariate aggregate choice data based on an approximation of latent cluster membership as a series of auxiliary conditionally Bernoulli random variables. Most recently, \cite{Klami2016} proposed a framework for Bayesian size-constrained microclustering using finite mixture models that uses a prior over the cluster sizes rather than sample membership in clusters. 

In contrast to previous work, here we formulate the problem of SCC as a Bayesian decision problem built on finite mixture models. Our decision theoretic approach is uniquely suited for cases in which the size-constraints reflect an extrinsic restriction related to the decision making processes rather than an internal restriction reflecting to the data generation process. In developing our decision theoretic framework we propose a novel loss function that combines the decision theoretic loss-functions introduced in \cite{Rastelli2016} for optimal cluster assignment with the Aitchison distance from Compositional Data Analysis \citep{Aitchison1992} to balance between assignment of observations into clusters to which they have high posterior probability of membership and satisfying the specified size constraints. In further contrast to previous methods, our approach allows for two different forms of size constraints on clusters. The first form of constraint consists of the label-switching sensitive specification of cluster sizes; here, size constraints are identified with a given group labeling. The second form of constraint consists of the label-switching invariant specification of cluster sizes; where target group sizes are given but not identified with a specific group labeling. This latter label-switching invariant form also ensures that cluster membership is robust to label-switching problems that can arise with mixture models \citep{Stephens2000}. Finally, by taking a fully Bayesian approach, we are able to integrate prior knowledge regarding cluster membership and characteristics beyond size-constraints, quantify the uncertainty in our measurements and inferences, and handle datasets with small sample size. 

Our motivation for developing this approach was, as previously introduced, to perform balanced clustering of individuals into teams for a Harry Potter \citep{Rowling1999} themed scavenger hunt. In this way, our goal was to implement a computational, rather than magical, version of the Sorting Hat from the Harry Potter book series. Like the literary Sorting Hat, we wanted to ensure that the resulting houses (clusters) maintained the house ``characteristics" described in the book series. To accomplish these goals we model each individual in our dataset as a mixture of the four Hogwarts houses. Our model jointly estimates how each of the four houses would respond to a set of categorical survey questions and the mixture weights representing each individual as a mixture of the four houses. In contrast to other implementations of the Sorting Hat currently available to muggles (a term for non-magical persons), our implementation is fully probabilistic.

The structure of this paper is as follows. Section \ref{general_theory} provides a background on finite mixture models and Bayesian decision theory. Section \ref{loss_function} introduces our proposed loss functions for SCC. Section \ref{categoricalMM} develops a mixture model for categorical survey data. Section \ref{posterior_inference} discusses our posterior simulation scheme and our methods for optimizing the proposed loss functions. Section \ref{identifying_clusters} introduces a relabeling algorithm to reidentify cluster labels which can be switched during posterior simulation. Section \ref{simulated_results} demonstrates the use of our categorical mixture model and decision theoretic approach on two simulated datasets. Finally section \ref{sortinghat_methods} applies our methods to a real dataset consisting of survey results from 21 scavenger hunt participants. 

\section{Methods} 
\label{methods}

In this section, we will develop a decision theoretic approach to size-constrained clustering (SCC) with finite mixture models. There are two key conceptual components of this approach. The first is the restatement of SCC as a Bayesian decision problem. The second component is the development of a loss function that balances between assignment of observations into clusters to which they have the highest posterior probability and satisfying the specified size constraint. In addition, we further develop a specific Bayesian finite mixture model for categorical survey data which we make use of in subsequent sections. 

\subsection{Mixture Models and Decision Theory}
\label{general_theory}
For a dataset $X$ consisting of $N$ univariate or multivariate observations, $(x_1,\dots, x_N)$, we describe the data as originating from a mixture of $K$ distinct distributions, $p(x_i | \phi_k)$ where $\phi_k$ is a parameter vector specific to the $k$-th distribution,  \(i=1,...,N\), and \(k=1,...,K\). These $K$ distributions represent the clusters in the dataset and for each observation $x_i$ we introduce a corresponding vector 
\[\theta_i : \left\{ (\theta_{i1}, \dots, \theta_{iK}) \in \mathcal{S}^K \right\} \] 
where $\mathcal{S}^K$ represents the $K$ dimensional simplex 
\[\mathcal{S}^K = \left\{  \left( x_1, \dots, x_N \right) \in \mathcal{R}^K | 
x_i > 0, i = 1,2,\dots, N; \sum_{i=1}^N x_i = k\right\} .\] 
$\theta_{ik}$ represents the proportion of observation $i$ that is characterized by cluster $k$, or equivalently the probability that observation $i$ belongs to cluster $k$. We write the likelihood of observation $i$ in terms of these clusters as 

\begin{equation}
p(x_i|\theta, \phi) = \sum_{j=1}^K\theta_j p(x_i|\phi_j). 
\end{equation}
Bayesian approaches to mixture models provide the posterior distribution $p(\theta, \phi | X)$ allowing us to quantify uncertainty in the cluster membership as well as in the parameters of the $K$ clusters.. 

Given the posterior distribution $p( \theta, \phi \vert X)$, we can formulate the task of cluster assignment as a decision problem. Let $a_i \in \{1, \dots, k\}$ denote the assigned cluster of observation $i$.  We want a cluster assignment $\hat{a} = \left( \hat{a}_1 , \dots, \hat{a}_N\right)$ that minimizes a specified loss function $\mathcal{L}(a,z)$.  Our loss function $\mathcal{L}(a,z) : \mathcal{Z}^{2\times N} \rightarrow \mathcal{R}$ quantifies the amount of loss we experience when we choose a cluster assignments $a$ when the true cluster assignments are $z$.

While $z$ is unknown, we may choose to minimize our expected loss given our posterior uncertainty in $z$. 
Thus, given $p(z|X)$, we want to find the assignment $\hat{a}$ which minimizes 
\[ \hat{a} = \argmin_{a \in \mathcal{Z}^N} \int \mathcal{L}(a,z) p(z | X) dz. \]
Note that given posterior samples from $p(\theta, \phi \vert X)$ we may obtain posterior samples from $p(z|X)$ by sampling $z_i \sim \text{Categorical}(\theta_i)$ for each observation $i$, implicitly marginalizing over the parameters $\theta$ and $\phi$. Given $T$ posterior samples from $p(z \vert X)$ we can approximate the above integral by 
\begin{equation} \label{optimization}
\hat{a} \approx \argmin_{a \in \mathcal{Z}^N} \frac{1}{T}\sum_t \mathcal{L}(a,z_t) p(z_t | X).
\end{equation} 
In what follows we describe our proposed loss function $\mathcal{L}(a,z)$ for SCC and develop a categorical mixture model for survey data thus specifying $p(z \vert X)$ for a target application.  

\subsection{Loss Function}
\label{loss_function}
The choice of loss function requires balancing the assignment of observations into the cluster of their highest predisposition and enforcing size constraints on the clusters.  For example, consider the problem of allocating two sandwiches (one Thai peanut chicken and one ham and swiss) to two people. Consider that person A has no particular preference while person B has a peanut allergy. In this case the decision making process should be driven by person B who has a strong preference against the Thai peanut chicken. To mimic this type of decision making behavior, we want our loss function to assign higher loss to the assignment of an observation A into a cluster $k$ to which it has a very low predisposition (low value of $\theta_{A,k}$) compared to the assignment of an observation B without strong preference ($\theta_{A1}\approx \dots \approx \theta_{Ak} \approx \dots \approx \theta_{A,K} \approx 1/K$) into that same cluster. In addition, we also want to the loss to increase as the cluster sizes deviate from the target size constraint. For example, we may want to penalize the choice to split the ham and swiss sandwich between person A and B while wasting the Thai peanut chicken sandwich. This second feature is the essential requirement that differentiates SCC from standard clustering. To achieve these dual goals we propose a new loss function that is the sum of two component functions. 

\subsubsection{Loss with respect to the true cluster assignments}
The first component function penalizes assignment of observations into clusters in which they have low posterior probability. Specifically, given the proposed cluster assignments for $N$ observations $a = (a_1, \dots, a_N)$, we want a loss function that penalizes deviations from the true cluster assignments $z=(z_1, \dots, z_N)$. While there are a number of potential loss functions that may serve our purposes, we follow the work of \cite{Rastelli2016} and choose the Variation of Information (VI) as it represents a proper distance metric over the space of partitions and is invariant under label-switching \cite{Meila2007}. While the first property is mathematically attractive, this second property is particularly important as it ensures that the action $\hat{a}$ is invariant to label switching problems that can occur in poorly identified mixture models \citep{Stephens2000}. 
 We denote the use of Variation of Information as a loss function in this way by
\[\mathcal{L}_{VI}(a,z) = 2H(a,z) - H(a) - H(z)\]
where $H(a,z)$ is the joint entropy of $a$ and $z$, $H(a)$ is the entropy of $a$, and $H(z)$ the entropy of $z$. Details on the calculation of  these quantities can be found in Appendix \ref{app_entropy_calculations}. 

\subsubsection{Loss with respect to the desired size constraints} 
The second component function penalizes deviation from the desired size constraints. 
To formalize this notion of size constraints, we must first introduce the operator $\mathcal{C} : \mathcal{Z}^N \rightarrow \mathcal{S}^K$ which maps an assignment vector $a$ to the composition of the groups under that assignment as
\begin{equation} \label{eq_closure}
\mathcal{C}(a) = \left( \frac{\sum_{j=1}^N I(a_j = 1)}{N}, \dots, \frac{\sum_{j=1}^N I(a_j = K)}{N}  \right)
\end{equation}
where $I(\cdot)$ denotes the indicator function. 

With this notation, we can introduce two different forms of size constraints on clusters. 
We call the first the label-switching sensitive size constraint, where group sizes are specified and identified with a specific cluster labeling. We may write this form of label-switching sensitive size constraint as a desired composition of the $K$ clusters as $\eta = (\eta_1, \dots, \eta_K)$ such that $\eta \in \mathcal{S}^K$. 
We call the second form the label-switching invariant size constraint, where group sizes are specified but are not identified with a specific cluster labeling. 
We may denote this form of label-switching invariant size constraint as $\eta_\sigma = (\eta_{\sigma(1)}, \dots, \eta_{\sigma(K)})$ where we let $\sigma = (\sigma(1), \dots, \sigma(K))$ denote a permutation of the cluster labels $\{1, \dots, K\}$. Finally, we also introduce the notation $\eta_{(\sigma)}$ as shorthand when a statement holds for $\eta$ or $\eta_\sigma$.

With this formulation of cluster size constraints we can now specify the second component of our loss function as a measure of the distance between the desired group composition $\eta_{(\sigma)}$ and our proposed group composition $\mathcal{C}(a)$. While there are a number of distance metrics that have been proposed to measure the distance between two compositional vectors \citep{Martin1999} we have chosen the Aitchison distance from Compositional Data Analysis \citep{Aitchison1992} as it is well studied and invariant to perturbation (translation in log-ratio space). As different size constraints can be formulated as perturbations of one another, this perturbation invariance ensures that the distance is consistent and well behaved under any size constraint \citep{Martin1999}. Importantly, perturbation invariance is relatively unique to the Aitchison distance and is not shared by other common distances/dissimilarities applied to compositions such as the Angular or the Bhattacharyya distances \citep{Martin1999}. 

The Aitchison distance, which we denote $d_A(x,y)$ is a measure of compositional distance between two vectors ${x,y} \in \mathcal{S}^D $ and is given by \citep{Aitchison1992} 
\[ d_A(x,y) = \sqrt{\frac{1}{2D} \sum_{i=1}^{D}\sum_{j=1}^D
\left(\ln\frac{x_i}{x_j}-\ln\frac{y_i}{y_j}\right)^2} . \]
Thus for the label-switching sensitive size constraint specification we may introduce $d_A(\mathcal{C}(a), \eta)$ as a measure of the distance between the desired size constraint and the proposed group sizes. 
For the label-switching invariant size constraint specification we may introduce $\min_\sigma \lbrace d_A(\eta_\sigma, \mathcal{C}(a))\rbrace$ as the measure of distance between the desired size constrained and the proposed group sizes. 

Since the Aitchison distance grows to infinity if any cluster decreases to zero, we introduce a small pseudo-count into our measurement of the relative abundances $\mathcal{C}(a)$ so that the distance is well behaved and does not dominate our overall loss function in the presence of zero counts. This slight modification allows groups to be empty if needed. 
We denote the pseudo-count augmented relative abundance vector 
\[ \mathcal{C}(a, \delta) = \left( \frac{\sum_{j=1}^N I(a_j = 1)+\delta}{N(1+\delta)}, \dots, \frac{\sum_{j=1}^N I(a_j = K)+\delta}{N(1+\delta)}  \right) \]
where $\delta$ should typically be a small positive constant in $(0, 1]$ \citep{egozcue2015_book}. 

\subsubsection{Combining the Variation of Information and the Aitchison Distance}
By combining $\mathcal{L}_{VI}$ with the Aitchison distance we may denote our proposed loss function for label-switching sensitive size constraints by 
\begin{equation} \label{lossfxn}
\mathcal{L}(a,z) = \mathcal{L}_{VI}(a,z) + \lambda \cdot d_A(\eta, \mathcal{C}(a,\delta))
\end{equation}
and for label-switching invariant size constraints as 
\begin{equation} \label{lossfxn_perminvariant}
\mathcal{L}(a,z) = \mathcal{L}_{VI}(a,z) + \lambda \cdot \min_\sigma \lbrace d_A(\eta_\sigma, \mathcal{C}(a, \delta))\rbrace
\end{equation}
where we have introduced the scalar $\lambda$ as a tuning parameter that allows the importance of the size-constraint in the overall loss function to be adjusted. 

To provide some intuition regarding the behavior of both Equations \eqref{lossfxn} and \eqref{lossfxn_perminvariant} we note that if $\delta = 0$ and $\lambda > 0$ then the loss function will not allow two clusters that are present in $\eta_{(\sigma)}$ to be merged in $\hat{a}$ as the Aitchison distance will become infinite when one group is empty. 
However, so long as $\delta > 0$, clusters can be merged together although such merging will be penalized by a factor that varies inversely to $\delta$. 
Note that this also provides a means by which an analyst may force clusters to merge. For example, if the original mixture model inference was carried out on $K$ groups, the analyst may choose a size constraint over $\alpha$ groups such that $\ 1\leq \alpha < K$ and $\eta_{(\sigma)} \in \mathcal{S}^\alpha$. In this situation, smaller values of $\delta$ will push some of the $K$ clusters together such that $\mathcal{C}(\hat{a}) \in \mathcal{S}^\alpha$.
Finally, we note that if $\lambda=0$ then our overall loss function reduces to the VI loss function.

It should be noted that the label-switching invariant form (Equation \eqref{lossfxn_perminvariant}) requires the extra computational step of finding the minimum Aitchison distance over permutation of group labels which grows in computational complexity as $K!$. 
However, in cases where $K$ is relatively small ($\approx < 7$) we find that the calculation of  $\mathcal{L}_{VI}$ dominates the computational complexity. 
If $\eta_{(\sigma)}$ is balanced (i.e. $\eta_{(\sigma)} = (1/K, \dots, 1/K)$ such that equal cluster sizes is desired), the loss functions in Equations \eqref{lossfxn} and \eqref{lossfxn_perminvariant} are equivalent and label-switching invariant. Thus for balanced size-constraints no extra minimization over permutations is required for label-switching invariance. 

\subsection{Categorical Mixture Model}
\label{categoricalMM}
In order to apply our decision theoretic approach to SCC to both simulated and real categorical survey data, we develop the following mixture model. The mixture model we develop is similar to the Latent Dirichlet Allocation model of \cite{Blei2003} as well as the Population Structure model with admixture of \cite{Pritchard2000}. 

Let $N$ represent the number of respondents to a survey of $Q$ questions. Let $V_q$ represent the number of possible responses to question $q \in \{1, \dots, Q\}$ and $K$ represent the number of clusters. 
To each of the $N$ observations we introduce the vector $\theta_n \in \mathcal{S}^K$ which represents the mixture weights for each observation in terms of the $K$ latent clusters. 
For each of the $Q$ questions, we introduce the vector $\phi_{k,q} \in \mathcal{S}^{V_q}$ which represents the probability that an observation from cluster $k$ would choose option $v_q$ for question $q$. 

We denote the survey responses as the $N\times Q$ matrix $X$ where $x_{nq} \in \{1,\dots, V_q \}$. With this notation we may write the likelihood of our data as 
\begin{align}
p(X|\theta, \phi) &= \prod_{n=1}^N\prod_{q=1}^Q \left( \sum_{k=1}^K p(x_{nq}|\phi_{kq}, \theta_{n})\right) \\
&=\prod_{n=1}^N\prod_{q=1}^Q\left(\sum_{k=1}^K\phi_{kqw_{nq}}\theta_{nk}\right)
\end{align}
In addition we specify Dirichlet priors for $\phi_{kq}$ and $\theta_n$ such that 
\begin{eqnarray}
\theta_n &\sim & \text{Dirichlet}(\alpha_n) \\
\phi_{kq} &\sim & \text{Dirichlet}(\beta_{kq})
\end{eqnarray}
where we introduce the hyper-parameters $\alpha_n \in \mathcal{R}^K$ and $\beta_{k,q} \in \mathcal{R}^{V_q}$ to reflect our prior information. 
By adopting the Dirichlet distribution for $\theta_n$ as opposed to the more generally flexible logistic-normal distribution we are implicitly assuming that the probabilities of a respondent belonging to each of the $K$ groups are independent except for the linear constraint of the simplex. This independence assumption is most evident when one considers that the Dirichlet distribution can be reparameterized as the normalization of independent gamma random variables. Our Dirichlet prior for $\phi_{kq}$ implies a similar independence assumption regarding the probabilities for a member of group $K$ responding to question $q$. While we believe that there is likely some covariation between components both in the $\theta_n$ and $\phi_{kq}$ vectors in our dataset, we believe this covariation is likely small and we make these simplifying assumption to reduce the number of parameters the must be inferred as our datasets are relatively small. This assumption could be relaxed by instead replacing the Dirichlet priors with Logistic-Normal priors. 

Finally, as described above, we model the distribution over the true cluster assignments $z$ given $\theta$ as 
\[ p(z_n \vert \theta_n) \sim \text{Categorical}(\theta_n)\]for an individual $n$. 

\subsection{Posterior Inference and Decision Theory Optimization}
\label{posterior_inference}
We perform posterior inference of $p(\theta, \phi, z | X)$ using the No-U-Turn-Sampler (NUTS) of \cite{Hoffman2014} that is provided in the Stan modeling language  \citep{Carpenter2017}. 
For the analysis of both our simulated and real datasets, four chains were run in parallel, for each chain we discarded 1000 samples as burn-in and then collected the subsequent 1000 samples as our representative sample of the posterior distribution. Sampler convergence was assessed by verifying that all sampled parameters had a modfied Gelman and Rubin statistic ($\hat{R}$; \cite{Gelman1992}) less than 1.01 and by manual inspection of sampler trace and autocorrelation plots. 

In order to preform the minimization specified in Equation \eqref{optimization}, we make use of the \pkg{rgenoud} package for R which uses a genetic optimization algorithm for mixed integer programing \citep{Walter2011}.  We found that the algorithms in the \pkg{rgenoud} package produced more stable results than the greedy algorithm of \cite{Rastelli2016} which we found was more sensitive to local optima. This optimization was preformed using custom R scripts with key functions compiled to C++ for computational efficiency using the functionality built into the Stan modeling language. 

All code including the mixture model inference and the decision theory optimization is provided in \ref{suppB}. 

\subsection{Identifying Clusters}
\label{identifying_clusters}
For label-switching invariant loss functions, the minimization of Equation \eqref{optimization} can lead to label switching in the Bayes action $\hat{a}$ compared to the labeling of $\theta$ or $z$. 
If the posterior for $\theta$ is identifiable, cluster labels for $\hat{a}$ can be recovered even when using a label-invariant loss function by finding the permutation of cluster labels that maximizes the posterior probability as follows. 
We will denote the elements of an assignment vector $\hat{a}$ as belonging to the permutation $\sigma$ of group labels $K$ such that 
$\hat{a}_i \in \left\{ \sigma(1), \dots, \sigma(K)\right\} $ where 
$\sigma(i) = j$ denotes that the group labeled by $i$ in the assignment vector $\hat{a}$ corresponds to the group labeled by $j$ in the identified posterior for $\theta$. 
Let $S$ denote the $K\times K$ matrix where
\[ s_{ij} = \sum_{t=1}^T\sum_{n=1}^N \log \theta^{(t)}_{nj} I(\hat{a}_n=i) \]
where $\theta^{(t)}$ denotes the $t$-th posterior sample of $\theta$.   
With this notation we can write the identified assignment vector as 
\[ \hat{a}^* = (\hat{\sigma}(a_1), \dots, \hat{\sigma}(a_N)) \]where 
\[ \hat{\sigma} = \argmax_{\sigma} \sum_{k=1}^K s_{\sigma(k) k}. \]
Thus the matrix $S$ essentially provides a look-up table that allows quick calculation of the posterior support for a given permutation of group labels in the assignment vector. This permutation is done to bring the group labels in the assignment vector into alignment with the group labels in an identified posterior. 

As with the calculation of the label-switching invariant loss function for unequal group size-constraints in Equation \eqref{lossfxn_perminvariant}, this optimization to identify groups grows as $K!$. However, as each step of the optimization simply involves a sum of $K$ values that can be quickly accessed from the matrix $S$, we find brute force maximization scales efficiently even for moderate sized $K$. 

\section{Results on Real and Simulated Data}
We demonstrate our decision theoretic approach to SCC on two simulated datasets and one real dataset of categorical survey responses. As our approach allows for both label-switching sensitive and label-switching invariant size constraint specifications as well as balanced and arbitrary unbalanced size-constraints, we developed two simulated datasets to demonstrate our proposed approach. The first simulated dataset consists of even group sizes and is useful in demonstrating balanced clustering using either the label-switching sensitive or label-switching invariant loss functions which are equivalent in the case of balanced clustering. The second simulated dataset consists of uneven group sizes and is useful in demonstrating an unbalanced size constraint either with a specified group labeling (label-switching sensitive) or no specified group labeling (label-switching invariant). Finally, we analyze the dataset which motivated the development of these approaches, the balanced clustering of scavenger hunt participants into four Harry Potter houses using categorical survey data.   

\subsection{Simulations} \label{simulated_results}
As proof of concept, we conducted two simulations with responses of 20 individuals ($N=20$) from 3 clusters ($K=3$) responding to 10 questions ($Q=10$) each with 3 possible answers ($V_q = 3$ for $q \in {1, \dots, Q}$). We chose these parameters so that all posterior distributions could be easily visualized with ternary diagrams. The first simulation contains balanced group sizes while the second contains unbalanced group sizes. Details regarding the simulation procedure can be found in \ref{suppA}.

For all analyses of simulated datasets, prior hyper-parameters $\beta$ were based off of the true simulated hyper-parameters with added uniform random noise (see \ref{suppB}) while prior hyper-parameters for $\alpha$ were set to $\alpha_n = (.5, .5, .5)$ for all $n \in \{1, \dots, N\}$ simulating no prior information other than that each individual was likely dominant in one of the three clusters. For both of our simulated datasets we obtained a posterior sample of $p(\theta, \phi, z|X)$ using MCMC as described in Section \ref{posterior_inference}.

\begin{figure} % figure 1
\includegraphics[width=5in]{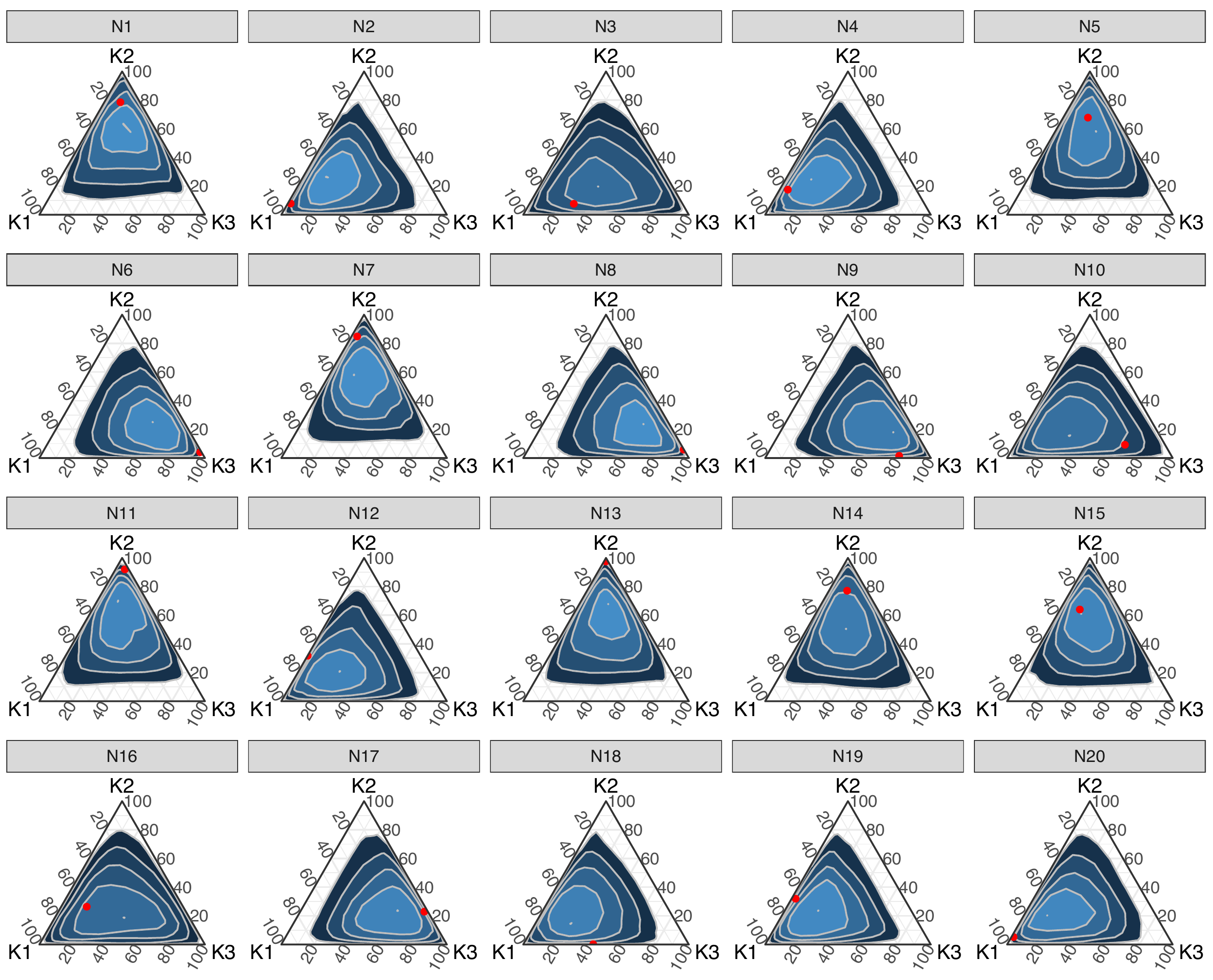}
%\vspace{6pc}
\caption[]{Posterior distributions for $\theta$ on simulated data with equal group sizes places regions of highest posterior probability near the true simulated values for $\theta$. Posterior distributions for the composition of 20 respondents (N1-N20) among 3 clusters (K1-K3) are shown using ternary diagrams with kernel density estimates. Kernel density estimates were calculated using the function \emph{stat\_density\_tern} from the R package \pkg{ggtern} \citep{Hamilton2017} with the parameter \emph{bins=5}. Red dots represents the true simulated value of $\theta$ for each respondent.} 
\label{sim_post_theta}
\end{figure}

The posterior distribution for $\theta$ for the simulated dataset with equal group sizes is shown in Figure \ref{sim_post_theta} along with the true values for $\theta$ (red circles). We see that for the majority of the simulated observations our posterior inference places the region of highest posterior probability near the true value for $\theta$. The posterior distribution for $\theta$ in the case of unequal group sizes and for $\phi$ in both the equal and unequal group size cases are shown in \ref{suppA} (Figures S1-S3). % XXX NOT HYPERLINKED...
All ternary diagrams were created using the R package \pkg{ggtern} \citep{Hamilton2017}. 
To prevent over-plotting, the functions \emph{geom\_density\_tern} or \emph{stat\_density\_tern} was used to calculate and plot kernel density estimates. 
In all cases we find that our posterior inference places the highest posterior probability near the true parameter values. 

In order to compare the performance of the VI loss function alone to our size-constrained loss function provided in Equations \eqref{lossfxn} and \eqref{lossfxn_perminvariant}, we preformed the minimization given in Equation \eqref{optimization} with the following parameter settings for the loss functions. 
For both simulated datasets we set $\delta = 0.1$ and we calculated $\hat{a}$ for the VI loss function alone by setting $\lambda=0$. Note that in this case \eqref{lossfxn} and \eqref{lossfxn_perminvariant} are identical. 
For the simulated dataset with even group sizes, we calculated the size-constrained loss function of Equation \eqref{lossfxn} with even target group sizes by setting $\lambda=1$  and $\eta =(1/3, 1/3, 1/3)$. 
For the simulated dataset with uneven group sizes, we calculated the size-constrained loss function of \eqref{lossfxn} using the known target cluster sizes for $\eta$ as well as the label-switching invariant loss function of \eqref{lossfxn_perminvariant} by using a random label permuted version of the true cluster sizes for $\eta$; in both cases we set $\lambda=1$. For all loss-function optimizations the function \emph{genoud} was run from the package \pkg{rgenoud} using the a population size of 3000, 20 wait generations, and 2000 maximum generations.

The results of our decision theory analysis is shown in Table \ref{decision_theory_table_sim}. 
We find that the inclusion of size-constraints in our loss function improves both the Variation of Information and accuracy of $\hat{a}$ (with respect to the true cluster assignments) compared to using the VI loss function alone.   

\begin{table}[t]
\centering
\caption{The performance the label-switching sensitive (LSS) and label-switching invariant (LSI) loss functions given in Equations \eqref{lossfxn} and \eqref{lossfxn_perminvariant} outperform the VI loss function of \cite{Rastelli2016} in terms of both accuracy and variation of information from the true cluster designations on two simulated dataset. The Even dataset consisted of 3 clusters with 7, 7, and 6 members respectively. The Uneven dataset consisted of 3 clusters with 8, 7, and 5 members respectively.}
\label{decision_theory_table_sim}
\begin{tabular}{c|ccc|cccc}
\cline{2-8}
\multicolumn{1}{l}{} & \multicolumn{3}{c}{Even} & \multicolumn{4}{c}{Uneven} \\ \hline
Observation & True Cluster & LSS & VI & True Cluster & LSS & LSI & VI \\ \hline
1 &   2 &   2 &   1 &   2 &   2 &   2 &   1 \\ 
2 &   1 &   1 &   1 &   1 &   1 &   1 &   1 \\ 
3 &   1 &   1 &   1 &   1 &   1 &   1 &   1 \\ 
4 &   1 &   1 &   1 &   1 &   1 &   1 &   1 \\ 
5 &   2 &   2 &   1 &   2 &   2 &   3 &   1 \\ 
6 &   3 &   3 &   1 &   1 &   1 &   1 &   1 \\ 
7 &   2 &   2 &   1 &   2 &   2 &   2 &   1 \\ 
8 &   3 &   3 &   1 &   3 &   3 &   3 &   1 \\ 
9 &   3 &   3 &   1 &   2 &   1 &   1 &   1 \\ 
10 &   3 &   3 &   1 &   3 &   3 &   3 &   1 \\ 
11 &   2 &   2 &   1 &   1 &   1 &   1 &   1 \\ 
12 &   1 &   1 &   1 &   1 &   3 &   1 &   1 \\ 
13 &   2 &   2 &   1 &   3 &   3 &   3 &   1 \\ 
14 &   2 &   2 &   1 &   1 &   1 &   1 &   1 \\ 
15 &   2 &   2 &   1 &   2 &   3 &   2 &   1 \\ 
16 &   1 &   3 &   1 &   3 &   3 &   3 &   1 \\ 
17 &   3 &   3 &   1 &   2 &   2 &   2 &   1 \\ 
18 &   3 &   1 &   1 &   1 &   3 &   3 &   1 \\ 
19 &   1 &   1 &   1 &   3 &   3 &   3 &   1 \\ 
20 &   1 &   1 &   1 &   2 &   2 &   2 &   1 \\  \hline
VI from True Cluster &  & 0.80 & 1.58 &  & 1.45 & 1.24 & 1.56 \\
Accuracy &  & 0.90 & 0.35 &  & 0.80 & 0.85 & 0.40 \\ \hline
\end{tabular}
\end{table}

\subsection{The Sorting Hat} \label{sortinghat_methods}
A convenience sample was obtained from 21 individuals, each of whom took a survey of multiple choice questions with variable numbers of choices for each question. Of the 8 questions on the survey, 7 were based on the Sorting Hat quiz available on the pottermore website \citep{Rowling} while 1 of the questions was of our own design. The survey questions and response options are given in \ref{suppA}. Each individual was provided with a URL that directed them to a Internet-based form where they could input their responses without seeing the responses of other respondents. The sample consisted of 10 males and 11 females, the majority of whom were either medical professionals or graduate students (18/21). The anonymized participant responses are given in \ref{suppB}. 

Posterior sampling and assessment of convergence was done as described in Section \ref{posterior_inference}. Prior information for $\phi$ was based on a descriptive analysis of 10,000 samples from the official Pottermore sorting hat quiz \citep{reddit_pottermore} and encoded in the hyper-parameters $\beta$. Our choices of  $\beta$ can be found in \ref{suppB} (Table S1). We assumed that we had no prior knowledge regarding which individuals belong to each house, however we assume that each individual is dominant in one of the four houses. We encode this prior assumption as $\alpha_n = (.3, .3, .3, .3)$ for all $n \in (1, \dots, N)$. 

\begin{figure} % figure 1
\includegraphics[width=5in]{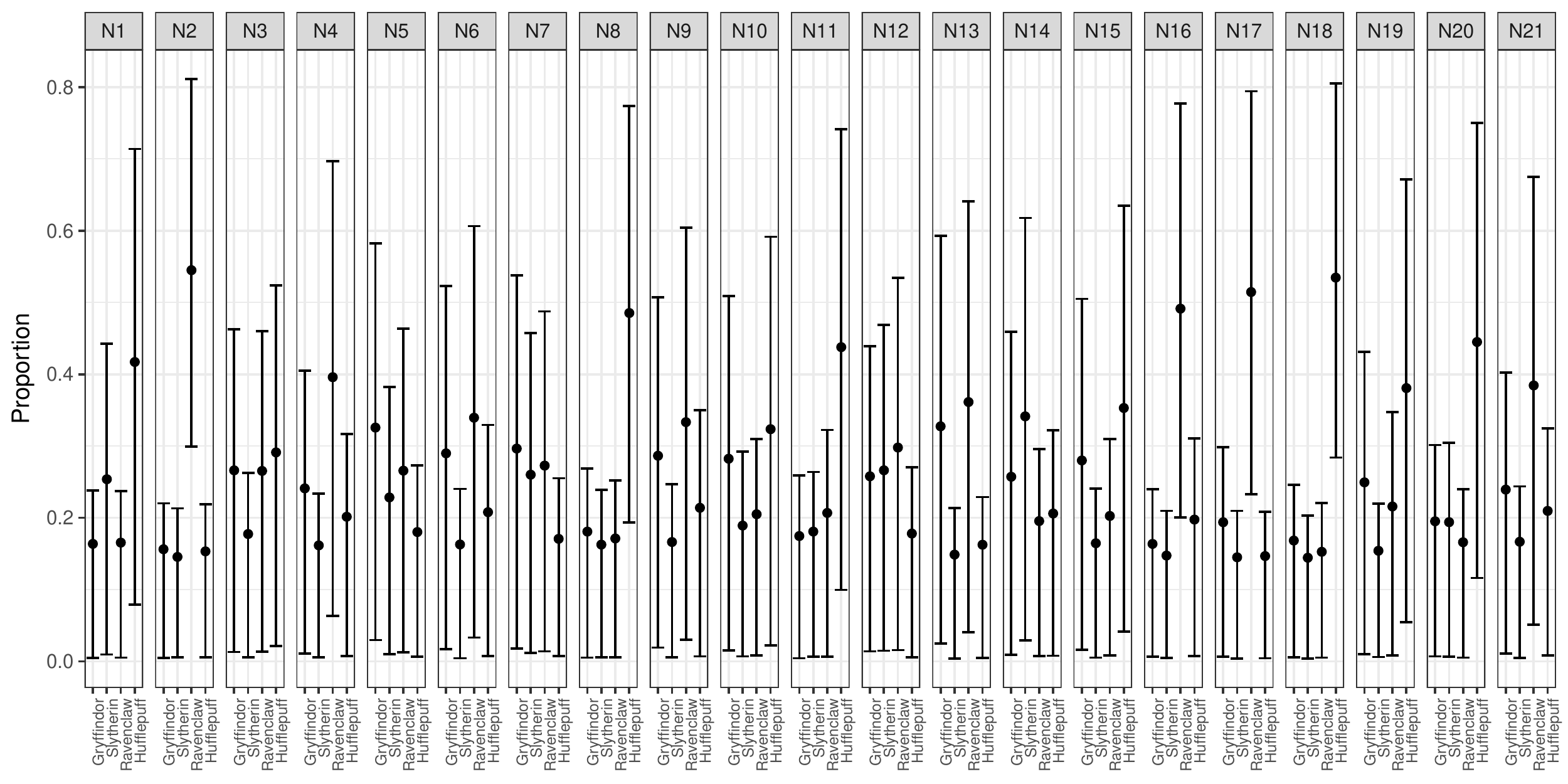}
%\vspace{6pc}
\caption[]{Mean and 95\% credible intervals of marginals of posterior distribution, $p(\theta|x)$, of house membership for each of the 21 respondents to the administered survey. As $K>4$ full joint distributions cannot be plotted as ternary diagrams. The majority of respondents were dominant in either Ravenclaw or Hufflepuff, whereas only one respondent (N14) was dominant in Slytherin and only two respondents (N5 and N7) were dominant in Gryffindor.}
\label{real_post_theta}
\end{figure}

The marginal posterior distribution for $\theta$ is shown in Figure \ref{real_post_theta}. We find that the majority of our respondents are dominantly Hufflepuff (9/21) and Ravenclaw (9/21) with only one individual dominant in Slytherin and two individuals dominant in Gryffindor. Our finding that the vast majority (86\%) of respondents were Ravenclaw and Hufflepuff dominant is in agreement with the work of \cite{Wilson2017}, who found that 87\% of approximately 500,000 American volunteers who completed a 21 question survey were Hufflepuff (30\%) and Ravenclaw (47\%). Similarly, \cite{Gouda2016} found that out of a sample of 478 medical students, post-graduate trainees, consultants and general practioners, 76\% were Hufflepuff and Ravenclaw, although their study only utilized a single question with four possible responses, each of which aligned with a single house. Finally, \cite{Crysel2015} examined a convience sample of 236 Harry Potter fans recruited from social media and found that only 61\% of the study participants self-reported being sorted by the Pottermore website sorting survey into Hufflepuff and Ravenclaw. Therefore, our finding that the majority of respondents were dominantly Hufflepuff and Ravenclaw is in good agreement with prior studies.

Having inferred the posterior distribution $p(\theta, \phi, z|x)$ we proceed to the decision theoretic component of our approach. As our goal was to push our clusters towards having equal sized groups, we used the label-switching sensitive form of our proposed loss function (Eq. \eqref{lossfxn}), set $\delta = 0.1$, and set $\eta = (1/4, 1/4, 1/4, 1/4)$. 
Again we compare the performance of our loss function against the the VI loss function proposed by \citep{Rastelli2016} alone. We calculated $\hat{a}$ for the VI loss function alone by setting $\lambda=0$, and for the size-constrained loss by setting $\lambda=1$. Loss function minimization was performed as detailed in Section \ref{posterior_inference}.

The results of our decision theory analysis are shown in Table \ref{decision_theory_table_real}. We find that the VI loss function alone clusters all respondents into a single house (Ravenclaw) whereas our size-constrained loss function groups respondents into approximately equal sized groups (5 in Ravenclaw Slytherin and Gryffindor  and 6 in Hufflepuff). This likely reflects the fact that there is poor separation of the clusters in the dataset. In this situation, the VI loss function alone does not capture our desire for balanced clusters and simply forces all individuals into the same house. In addition, as we would want from our proposed loss function, we find that respondents that were the most dominant in a given house appear to drive the ultimate Bayes action whereas respondents who were relatively even in all houses filled up the less common houses (e.g., Slytherin and Gryffindor). For example, we see that respondents N2, N16, N17 are sorted into Ravenclaw and Respondents N8 and N18 are sorted into Hufflepuff as would be expected given their dominance in the associated houses in Figure \ref{real_post_theta}. Whereas we see respondents N3, N10, and N19 are sorted into Gryffindor and N5, N6, N7, and N12 are sorted into Slytherin in order to satisfy the size-constraint as these respondents did not show any particular dominance in one house over another. Thus, we find that our size-constrained loss-function has produced equal group sizes while minimizing the degree to which respondents with strong asymmetry in their posterior distribution are misclassified. 

\begin{table}[t]
\centering
\caption{The label-switching sensitive (LSS) loss function of Eq. \eqref{lossfxn} is able to capture our desire for balanced housing (cluster) assignments whereas the VI loss function of \cite{Rastelli2016} alone cannot. Housing assignments under either the LSS loss function or the VI loss function alone are shown for each respondent to the 8 question survey sent to scavenger hunt participants.}
\label{decision_theory_table_real}
\begin{tabular}{c|cc}
\hline
Respondent & LSS         & VI        \\ \hline
1 & Hufflepuff & Ravenclaw \\ 
2 & Ravenclaw & Ravenclaw \\ 
3 & Gryffindor & Ravenclaw \\ 
4 & Gryffindor & Ravenclaw \\ 
5 & Slytherin & Ravenclaw \\ 
6 & Slytherin & Ravenclaw \\ 
7 & Slytherin & Ravenclaw \\ 
8 & Hufflepuff & Ravenclaw \\ 
9 & Ravenclaw & Ravenclaw \\ 
10 & Gryffindor & Ravenclaw \\ 
11 & Hufflepuff & Ravenclaw \\ 
12 & Slytherin & Ravenclaw \\ 
13 & Gryffindor & Ravenclaw \\ 
14 & Slytherin & Ravenclaw \\ 
15 & Hufflepuff & Ravenclaw \\ 
16 & Ravenclaw & Ravenclaw \\ 
17 & Ravenclaw & Ravenclaw \\ 
18 & Hufflepuff & Ravenclaw \\ 
19 & Gryffindor & Ravenclaw \\ 
20 & Hufflepuff & Ravenclaw \\ 
21 & Ravenclaw & Ravenclaw \\ \hline
\end{tabular}
\end{table}

\section{Conclusions}

We have proposed a method for introducing size constraints into finite mixture model based clustering through a decision theoretic approach. Our approach utilizes a novel loss function that is the combination of the variation of information \citep{Meila2007} and the Aitchison distance \citep{Aitchison1992} between the desired constraints and the proposed action. In addition, Our approach allows for arbitrary size constraints that can be either label-switching sensitive or invariant depending on the desired application. To demonstrate our approach, we have developed a mixture-model for categorical survey data that closely resembles both the Latent Dirichlet Allocation  model of \cite{Blei2003} and the Admixture Population structure model of \cite{Pritchard2000}. We applied this method to two simulated datasets, one with even group sizes and one with uneven group sizes, as well as one real-world dataset consisting of questions intended to sort survey respondents into Harry Potter houses. Our results demonstrate that a variety of size constraints can be used including clustering into balanced clusters, uneven clustering where specific group sizes are known and identified with specific cluster labels, and uneven clusters where group sizes are known but are not identified with specific cluster labels.

The improved clustering accuracy we see with our proposed loss function compared to the VI alone is not surprising as our loss function allows extra information on cluster sizes to be accounted for in the decision theoretic. When the clusters are well separated in the posterior it is likely that our approach will show only marginal improvements in terms of accuracy over the VI alone. However, when the clusters are poorly separated in the posterior as is the case in both our simulated and real datasets (Figures \ref{sim_post_theta} and \ref{real_post_theta}), we see that this extra information regarding cluster sizes produces a large increase in accuracy. As we expect many real world datasets to display poor separation in clusters, we believe our approach may be particularly appealing. 

Of course, accuracy alone is not always the primary goal of an analysis. For example, our application of SCC for sorting participants into Harry Potter houses is an example of a size constraint related to the decision making process and not the data generation process. Among previously published methods for SCC our approach is unique in taking a decision theoretic approach. However, our approach does not exclude cases in which size constraints occur in the data generation and decision making processes. In such cases, our decision theoretic approach can be build on a generative Bayesian mixture model that also models the size constraint through either a prior or likelihood model such the model of \cite{Klami2016}. 

While sorting individuals into different Harry Potter houses is a niche statistical challenge, SCC is a common problem that shows up in a variety of statistical arenas. While not an comprehensive list, we categorize potential applications of SCC into two groups: cases with fixed size constraints and loose size constraints. With fixed size constraints, it is essential that the resulting cluster sizes exactly match the desired constraint. Such situations may arise with adaptive clinical trial designs where treatment groups are blinded but aggregate totals are known \cite{Teel2015}, in market research when inferring individual purchasing habits given relevant covariate and aggregate purchasing patterns for a region, or if voting patterns were to be inferred given district totals and relevant covariates \citep{Teel2015}. In cases with fixed size constraints a small $\delta$ and large $\lambda$ value in Equations \ref{lossfxn} or \ref{lossfxn_perminvariant} ensure that the loss function is dominated by the size constraint thus ensuring the Bayes action corresponds to the required size constraint. With loose size constraints, desired cluster sizes are viewed as a guideline but exact adherence is not required. Such situations may occur when there is either weak prior information regarding true cluster sizes or there is some flexibility in group sizes centered around a desired constraint. The latter case is likely to be seen in industry applications where resources or supply is limited with some flexibility in its allocation or when planning wedding tables for a venue with fixed size tables and some flexibility in the density of chairs around the tables. In cases with loose size constraints, a large $\delta$ and small $\lambda$ ensure that clustering is driven largely by the true underlying group structure and only weakly by the specified size constraints. 

While our results demonstrate that our approach enables SCC through a decision theoretic approach, a number of challenges still remain. Chief among these challenges is the difficulty posed by optimizing functions over discrete spaces with potential local optima. Much of this difficult reflects the fact that integer programming is an NP-hard problem. Here we have adopted the genetic algorithm approach to integer programming provided by the \pkg{rgenoud} package \citep{Walter2011}. While we find this approach works well for our applications it can be computationally intensive and it is unclear how well this approach will scale to much larger datasets. However, we do note that the optimization of Equation \eqref{optimization} done for each of the datasets in this work required only about 10 minutes each on a single computer core, while the \pkg{rgenoud} package does support multi-core and multi-host parallelization. In this way parallelization could provide a route to expanding to larger datasets than we analyzed here. However, ultimately processing massive datasets in this manner, as would be needed for de-anonymizing voting records or similar sized problems, would likely require better integer programing algorithms. We also note that both the VI loss as well as our size-constrained loss function are susceptible to local optima. In our experience, local optima become more severe as the desired group sizes become more uneven and the overall optimization becomes more asymmetric; both balanced clustering and the label-switching invariant form of the size-constrained loss function alleviate some of the challenges of local optima by making the overall optimization symmetric with respect to label switching. Despite these challenges, we believe our proposed methods represent a useful new approach to SCC. 

Finally, despite the extensive view into the world of Harry Potter provided by J.K. Rowling, the mechanisms of magic remain a mystery. We believe that our findings shed new light on the inner workings of this magic. The severe house imbalance shown by our method within our population is also echoed in the work of \cite{Wilson2017}, \cite{Gouda2016}, and \cite{Crysel2015}. In the absence of size constraints and assuming that the wizarding community matches the studied populations in terms of house tendencies, we would expect overcrowding in the Ravenclaw and Hufflepuff common rooms whereas we expect that the Slytherin and Gryffindor houses would not be able to assemble Quidditch teams with such limited membership. Therefore, we conclude that either the wizarding community is more homogenous and balanced than the muggle (non-magical) populations studied to date or the magical sorting hat involves some size-constraining in its decision making process. In either case, we believe our results shed new light on the inner workings of the wizarding world.   

\appendix

\section{Variation of Information Calculations}\label{app_entropy_calculations}

We consider the calculation of Variation of Information \citep{Meila2007} by making use of the closure operation defined in Equation \eqref{eq_closure}. In keeping with the notation of \cite{Rastelli2016}, for two assignments $a$ and $z$ with $K_a$ and $K_z$ groups respectively we define the contingency matrix  as the $K_a \times K_z$ matrix with entries given by

\[n_{gh}^{az}  = \sum^N_{i=1}I(a_i=g)I(z_i=h)\]
where $g$ is an index for groups in $a$ and $h$ for groups in $z$ respectively.  Let $n_g^a$ and $n_h^z$ represent the sizes of groups $g$ and $h$ such that 
\[ n_g^a = \sum_{h=1}^{K_z} n_{gh}^{az}, \quad\quad n_h^z = \sum_{g=1}^{K_a} n_{gh}^{az}.\]
We define the entropies of $a$ and $z$ as 
\[ H(a) = -\sum_{g=1}^{K_a} \frac{n_g^a}{N}\log_2\frac{n_g^a}{N}, \qquad  H(z) = -\sum_{h=1}^{K_z} \frac{n_h^z}{N}\log_2\frac{n_h^z}{N} \]
and the joint entropy of $a$ and $z$ as 
\[ H(a,z) = -\sum_{g=1}^{K_a}\sum_{h=1}^{K_z}\frac{n_{gh}^{az}}{N}\log_2\frac{n_{gh}^{az}}{N}. \]
We can now define the Variation of Information loss as 
\[\mathcal{L}_{VI}(a,z) = 2H(a,z) - H(a) - H(z). \]

\section*{Acknowledgements}
We thank Sayan Mukherjee for his helpful and insightful comments as well as our friends and family who have endured our statistical antics. JS was supported in part by the Duke University Medical Scientist Training Program (GM007171).

% See instructions for supplement here: http://www.e-publications.org/ims/support/ims-instructions.html
\begin{supplement}[id=suppA]
\sname{Supplement A}
\stitle{Prior Specifications, Simulation Methods and Results, and Survey Questions }
\slink[doi]{} % Filled out by typesetter
\sdatatype{.pdf}
\sdescription{Prior specifications for both the simulated datasets and the Harry Potter questionnaire. Also contains description of methods for data simulation and results as well as the questions and possible responses for the questionnaire administered to scavenger hunt participants.}
\end{supplement}

\begin{supplement}[id=suppB]
\sname{Supplement B}
\stitle{Reproducible Code and Data}
\sdescription{Rmarkdown files and anonymized data necessary to reproduce results.}
\slink[doi]{} % Filled out by typesetter
\slink[url]{https://github.com/jsilve24/BaysianSortingHat}
\sdatatype{.zip}
\end{supplement}

% Begin Bibliography
\bibliographystyle{imsart-nameyear}
\bibliography{sorting_hat}

\end{document}